Performance of hybrid angle-energy dispersive X-ray diffraction and fluorescence portable system for non-invasive surface-mineral identification in Archaeometry.

Ariadna Mendoza Cuevas [1,†] and Jorge Fernández-de-Cossio Dorta-Duque[1,2]

1  Archaeometry Laboratory at Havana Historian's Office, San Ignacio entre Obispo y Aguiar Habana Vieja, CP 10 100 Havana, Cuba
2  Center for Genetic Engineering and Biotechnology, CP 11300 Havana, Cuba

†Corresponding author: ariadna@patrimonio.ohc.cu, ari.mendoz0@gmail.com

Abstract

Low power energy dispersive XRD-XRF portable instruments equipped with multiple angle scanning can take advantage of the shorter acquisition time of EDXRD with respect to ADXRD, and bring closer higher accuracy and resolution of inter-planar distance with those obtained by ADXRD. The data produced by this new hybrid configuration is correlated in the sense that a single XRF or XRD specimen appear in multiple spectra (the later shifted in energy for differing angles). Hence, for fully benefit from the richer data released by this configuration, the analysis should not be confined to the independent processing of the spectra, specialized hybrid data processing should be conceived. We previously reported some advances in the processing of the resulting 3D data (intensity, energy and angle). Here the analytical performance of the first hybrid angle-energy dispersive X-ray diffraction and fluorescence portable system is assessed for non-invasive surface mineral analysis of samples relevant for archaeometrical applications. We evaluate the performance on standard reference material and probe applicability of the methods so developed to identify stones (jadeite and omphacite), and pigments (Prussian blue) in the pictorial layer of modeled paintings. Discussion emphasize the improvement in accuracy of interplanar distance with respect to EDXRD taken at a single fixed angles, evaluate the resolution of AD/EDXRD data, and total acquisition time.

**Keywords:** XRF, EDXRD, portable XRD XRF, non-invasive analysis, Archaeometry.

I. Introduction

Angle Dispersive X-ray Diffraction (ADXRD) is the detection method most widely found in XRD instruments. ADXRD use a monochromatic energy beam to excite the sample, and the full diffractogram is obtained by measuring at multiple scattering angles. In contrast, Energy Dispersive X-ray Diffraction (EDXRD) irradiate the sample simultaneously with a broad energy range, and a single scattering angle is required to obtain a relative wide diffractogram (parallel acquisition). Consequently, faster acquisition time is reached by EDXRD with respect to ADXRD. Additionally, XRF lines are naturally detected with EDXRD setup. The isotropic nature of XRF and angular dependence of XRD can be used to distinguish each other's by the addition of a goniometer [1], permitting also to sort interfering XRF lines. An alternative method, used in back reflection EDXRD geometry to avoid insensitivity to sample morphology, separates diffraction and fluorescence peaks by tuning the excitation energy of an X-ray tube source to just below the absorption edge of the interesting elements [2].

The potential power of Portable (AD) XRD system in the analysis of art and archaeological objects (Archaeometry field) and the advantages and disadvantages of current instrument configurations were recently reviewed [3]. There are some indicated requirements worth to focus: accurate positioning of the instrument head with respect to the object [8]; reduction of acquisition time to reach lower $2\theta$ angle, larger $2\theta$ range and higher penetration of X-rays which may reach deeper layers; and finally to obtain a compact and light weight design.

The first report on EDXRD portable system for non-invasive analysis of cultural heritage [4] used a polychromatic beam from a low power (Cu or W anode, operated at 35 kV, 1 mA) X-ray tube, a Si-PIN detector and an arc shaped goniometer to sort XRF lines. This work shown the advantages of EDXRD, detecting two XRD lines (for quartz and pigments) in a short measuring time, typically 100 s, with a light weight instrument head (2 kg). However, in this setup, the characteristic lines from the chosen anode (Cu-k or W-L line) of the low power X-ray tube intrude in the middle of the energy range in the diffractogram, as an unavoidable instrumental XRF interference hindering the identification of specific pigments. In a second report by the same authors a combined system works independently in ADXRD (monochromatic Cu-K or W-L) and EDXRD (polychromatic W-L lines) was tested [5]. XRF and complementary ADXRD analysis were used for pigment identification in the painted statue of ''Tamonten holding a stupa'' from the Heian Period (794–1192). Meanwhile, by a comparison of the EDXRF spectra taken at 52° and the EDXRD-XRF taken at 10°, concluded that Au and Cu are rich in the surface layer, but Fe is rich in the underlying layer of the pedestal of the statue. Even when this work recognized that EDXRD is rather efficient in getting promptly the structural information on a specimen, they also pointed out that the line width of the diffraction pattern thus obtained is broader, especially in the wide $2\theta$ range, compared with a conventional ADXRD instrument. The mentioned system turned eventually into XRDF, an ADXRD based commercial system [3].

The main advantage of EDXRD systems comes from the high energy components of the white X-ray beam. Taking into account the highest energy limit of the available X-ray tubes, a different low power EDXRD-XRF portable system setup was proposed by the author [6]. The main distinctions are: 1) higher energy polychromatic beam (i.e. Pd or Ag anode) which favors EDXRD intensities in the measurable energy range, clean from instrumental XRF interference; 2) a more compact instrument by using a close/open goniometer which also improved flexibility for the analysis of complex geometry object; 3) existing portable XRF systems can be adapted to enrich XRD capabilities. Motorized multiple angle scanning were later incorporated in two prototypes based on this setup [7]. Individual EDXRD diffractograms of quartz standard and archaeometrical samples were evaluated [6,7] (12 lines detected, 300 s, operated at 28 kV, 200 µA [7]), and the full picture of unprocessed multiple angle spectra are represented by 3D Energy vs Angle plots, showing the strongest iso-d curves.

EDXRD at a single angle covers only a partial range of XRD lines, since small d-line values are centered in the energy range at higher angles, while larger d-line values are so located at lower angles. Hence, with multiple angle measurements a wider range of d values are spanned. Further, since XRD lines shift with angle, interferences with XRF lines at one angle can be sorted in another angle.

The data obtained with hybrid AD-EDXRD are highly correlated in the sense that the same XRF or XRD specimen usually appears in multiple spectra, the later shifted in energy according to the angle. To fully benefit from the richer data released by this hybrid system, specialized hybrid data processing should be conceived. Recently, we reported some methods for hybrid angle /energy dispersive XRD-XRF data processing [8] take advantage of this hybrid system scenario by analyzing the 3D (Energy, Angle and Intensity) information so generated. This method constructs a hybrid diffractogram that resume the correlated information for each specimen. A depth scanning based method for object positioning relevant for non-invasive analysis in Archaeometry were also advanced in that report, and the solution of the XRF interference is displayed through the lineal density plots E vs. Angle where XRF lines are more contrasted with respect to XRD [8]. These methods were evaluated with the 3D data of quartz standard collected with a low power AD EDXRD/XRF instrument using the multiple angle scanning EDXRD setup previously developed [7].

Applications of interest in Archaeometry are more challenging than evaluating powder standards, for example non-instrumental XRF interference are more common, and contaminating materials and layers complicate the analysis in actual non-invasive in situ applications. The recently developed instruments and methods has not been approached as hybrid system for those applications. In this manuscript we assess the performance of the hybrid angle-energy dispersive XRD and XRF portable system for non-invasive surface-mineral identification in Archaeometry. Our present specific purposes are: (1) to evaluate the proposed low power hybrid XRD-XRF system (instrument + data processing methods/software) to identify minerals, emphasizing improvement of d values accuracy with respect to EDXRD taken at a single angle, and to compare d-spacing resolution of EDXRD with ADXRD. This will be evaluated with a standard reference material, in mineral identification of green stone, which is useful to study archaeological objects presumably made on jade that are interesting for provenance and migration route determination. (2) To evaluate the possibility of hybrid to discriminate XRD signal of sample surface layer from underlayer signals. This will be evaluated in the identification of pigments in the pictorial layer of modeled paintings, not discernible using XRF alone. (3) To evaluate the total acquisition time reduction in mentioned applications supporting the time efficiency of EDXRD system.

## 2. Materials and methods

EDXRD-XRF diffractograms are obtained in an energy range from 0 to 22 keV at 28 kV and 200 µA using bremsstrahlung radiation emerging from an Ag anode tube (10–50 kV, 200 µA, 400 µm anode spot). A 3D data (Intensity (I), Angle ($\theta$), and Energy (E)) is collected using a Si-drift detector (SDD) (136 eV FWHM at Mn k$\alpha$). The SDD with a multichannel analyzer (MCA) was calibrated against several pure oxide (Si–$SiO_2$) and metal foils (Si, Ti, Cu, Pb, Zr, Ag). A linear calibration of channel numbers to these reference energies accurately fits within typical uncertainties of 0.09% about 4.51 keV and about 15.76 keV, with FWHM of 0.1933 eV and 0.2824 eV respectively.

Reproducible positioning of the measurement head with respect to a specific analyzed region of the object's surface (X and Y planes) is carried out by using a position-sensitive laser that measures the distance with an accuracy of 0.02 mm. The mechanical design of close/open goniometers allows to sweep a wide range of angles between the X-ray tube and detectors arms ($0 < \theta < 60$). This is particularly convenient for the analysis of objects with complex geometry, as is typical in cultural heritage and for increasing the d ranges covered.

EDXRD-XRF spectra from the studied samples were acquired at 15° along depth scanning (Z axis) measurements for optimum object position determination [8]. For this acquisition the sample were translated with respect to the measurement head (X-ray tube and detector) in a direction bisecting the angle formed by the beams. X-ray tube – sample distances (approx. 5 cm) were chosen to accommodate the samples.

The hybrid diffractogram are built by accumulating the counts along the iso-d curve from the measured 3D data with the background subtracted as described in (Mendoza Cuevas & Fernández-de-Cossio Dorta-Duque, 2016). The hybrid construction method and software have been improved here by pondering the peak contribution relative to signal/noise and the intensity, visibly gaining in quality and precision with respect to our previously reported construction. The XRF signals are identified from candidate specimens, and the intensity from each spectra are summed along the peak range to obtain a hybrid XRF spectrum.

A pelletized powder sample of the standard reference material $LaB_6$ certified by the NIST (Freiman & Trahey, 2000) was analyzed because its suitability for powder diffraction calibration of line position and shape in order to evaluate the accuracy of d–spacing values achieved and the possibility to reduce acquisition time without appreciable degradation of XRD signal. Measurements of standard $LaB_6$ were performed for 200 and 400 sec/angle at a maximum of 25 angles distributed uniformly in the logarithmic scale from 5° to 50°, as previously proposed (Mendoza Cuevas & Fernández-de-Cossio Dorta-Duque, 2016).

Stone mineral fragment of clear green jadeite from Guatemala, dark green jadeite and dark green omphacite were non-invasively analyzed to identify all XRD lines detected in the measured angular range. Measurements of green stone were performed between 50 –400 sec/angle at a maximum of 22 angles logarithmically distributed from 5° to 50°.

Model samples of oil painting on canvas prepared with zinc white and inert materials such as calcite and barite were analyzed. Its pictorials layers are composed of pure pigment Prussian blue (powder), usually not identified by XRF. Measurements of pigments were performed for 50 - 400 sec/angle at a maximum of 24 angles logarithmically distributed from 5° to 25°.

In order to evaluate the reduction of total acquisition time without appreciable degradation of signal, subset of the original measured hybrid 3D data were chosen with different number of angles and measurement time/angle. The logarithmic scale was procured in the selection of six angles taking into account the angular range of the interesting d-spacing values to cover, and to avoid the non-instrumental XRF lines of the analyzed sample.

## 3. Results and discussion

The performance of the hybrid low power X-ray diffraction and fluorescence portable system for surface mineral identification in Archaeometry is evaluated with 3D-constructions and hybrid diffractogram. 3D-constructions are represented by I - Angle vs. Energy density-plots with logarithmically transformed axes [8]. Accuracy and resolution of d-spacing value is evaluated with a standard reference material, and acquisition times for each data are indicated.

### 3.1 Accuracy and resolution of d-spacing estimation

Figure 1 show some plots for the hybrid analysis of the standard $LaB_6$ on each the three datasets: 1) 25 diffractograms measured at angles uniformly distributed in the log scale, 2) six diffractograms included in the first dataset keeping a log scale distribution, and 3) single ED diffractograms measured at 12.23° and 23.84°. The log transformed density plot of the full multiple angle scanning dataset show the prominent d-spacing lines along with other weaker lines (Figure 1A). The higher values (4.15 Å) are not covered at 23.84°, and the lower values (< 1.2 Å ) are not covered at 12.23° (Figure 1C). Hence, by introducing multiple angle scanning capabilities a wider d-spacing range is spanned, delivering in turn additional counting statistic XRD line since each XRD and XRF lines typically appear in more than one angle. Accounting for this redundancy or correlation in the hybrid construction permit to improve accuracy of d-spacing values. Note that for the same angular range scanned in the ADXRD, a wider d range is covered in the hybrid system. Signals lower than 1 Å are observed in the Angle vs Energy density plot and the individual diffractogram.

The performance of the hybrid processing in terms of d-spacing accuracy is resumed in Table 1 and Figure 1D for 25 and six angles datasets. The theoretical d-spacing were calculated from the unit cell dimensions of the standard $LaB_6$ certificated by the NIST (Freiman & Trahey, 2000). The observed d-spacing were summarized with the estimated centroid of the computed hybrid diffractogram of each detected peak and the respective standard deviation. The difference of observed and theoretical d (error) are also listed in the table.

The errors of estimated d-spacing lay within one standard deviation, with the only exceptions of two in 38 cases, which lay well within two standard deviations. The errors with the 25 angles dataset vary in the order of thousandths, and are smaller in most of the cases when compared with the six angles dataset. The maximum error is 0.001589 in the former and 0.002062 in the later. This is consistent with the $\sqrt{N}$ rule of thumb improvements in accuracy, considering that resolution, intensity and quality of d-spacing peaks vary with angle. The six angles hybrid show no appreciable deterioration of XRD signals when compared with the 25 angles hybrid. Accuracy of d-spacing values could be improved with other selection of six angles if relevant prior knowledge of the sample is available.

The precision of the estimated d-spacing are appreciated by the fine peaks exposed by the hybrid diffractograms. However, by the way the hybrid is currently constructed, resolution cannot be directly inferred from the hybrid's peak width.

Resolution of ADXRD portable systems, conventionally estimated for $2\theta$ in the range 20° - 30° using an angular step of 0.1°, have been reported in the range 0.2° to 0.3° on average [3], which corresponds with mean d spacing resolution of 0.04205 Å. To attempt a comparison we chose the resolution of lines 2.939 Å and 4.15692 Å of $LaB_6$, laying in the extremes of the corresponding d-spacing range. The resolution of 0.1033 Å and 0.224528 Å was obtained by those lines respectively, with mean resolution of 0.0425 Å which is 3.8568 times the mean reported by ADXRD setup.

Table 1: Values of d, accuracy (error) and precision estimation (standard deviation) of the hybrid system on the measured standard $LaB_6$.

| hkl | d theo | d-obs (centroid) | error (25 angles) | error (6 angles) | Standard deviation | peak max |
|---|---|---|---|---|---|---|
| _001 | 4.15692 | 4.15704 | -0.00013 | 0.00748 | 0.074624 | 446.865 |
| _011 | 2.93938 | 2.93798 | 0.001399 | 0.001952 | 0.023719 | 1290.61 |
| _111 | 2.4 | 2.40133 | -0.00133 | -0.00077 | 0.014569 | 639.864 |
| _002 | 2.07846 | 2.07826 | 0.000202 | 3.8E-05 | 0.012303 | 319.465 |
| _012 | 1.85903 | 1.86073 | -0.0017 | -0.00129 | 0.007994 | 684.621 |
| _112 | 1.69705 | 1.69884 | -0.00178 | -0.00284 | 0.007492 | 301.858 |
| _022 | 1.46969 | 1.4719 | -0.0022 | -0.00318 | 0.00934 | 88.9519 |
| _122 | 1.38564 | 1.38613 | -0.00049 | -0.00188 | 0.005714 | 305.527 |
| _013 | 1.31453 | 1.31558 | -0.00105 | -0.00178 | 0.006689 | 191.804 |
| _113 | 1.25336 | 1.25575 | -0.00239 | -0.00371 | 0.006853 | 89.6139 |
| _222 | 1.2 | | | | | |
| _023 | 1.15292 | 1.15354 | -0.00062 | | 0.007829 | 41.8537 |
| _123 | 1.11098 | 1.11569 | -0.00471 | -0.00603 | 0.005083 | 91.6252 |
| _004 | 1.03923 | 1.04021 | -0.00098 | -0.0048 | 0.007534 | 16.6536 |
| _223 | 1.0082 | 0.999488 | 0.008713 | 0.008353 | 0.004449 | 46.7537 |
| _114 | 0.979795 | | | | | |
| _133 | 0.953662 | | | | | |
| _024 | 0.929515 | | | | | |
| _124 | 0.907113 | 0.908868 | -0.00175 | -0.00239 | 0.005438 | 31.886 |
| _233 | 0.886258 | | | 0.016673 | 0.00427 | 6.53558 |
| _224 | 0.848527 | 0.848083 | 0.000444 | | 0.006009 | 9.1882 |
| _034 | 0.831383 | | | | | |
| _134 | 0.815238 | 0.815997 | -0.00076 | -0.00275 | 0.004366 | 25.0314 |
| _333 | 0.799999 | | | | | |
| _234 | 0.77192 | 0.770857 | 0.001063 | -0.00195 | 0.004506 | 14.5845 |
| _225 | 0.723626 | | | -0.00437 | 0.002861 | 4.71978 |
| _334 | 0.712905 | 0.712234 | 0.000671 | -0.00141 | 0.00597 | 10.033 |
| _244 | 0.692819 | 0.690958 | 0.001861 | | 0.003128 | 8.23955 |
| _344 | 0.649201 | 0.648725 | 0.000476 | | 0.004401 | 7.4773 |
| _245 | 0.619676 | 0.620634 | -0.00096 | | 0.003732 | 4.74836 |
| _136 | 0.612904 | 0.608308 | 0.004596 | 0.004719 | 0.002623 | 5.17991 |
| _345 | 0.587877 | 0.588965 | -0.00109 | | 0.004166 | 3.96167 |
| _046 | 0.576461 | 0.57638 | 8.07E-05 | | 0.002716 | 3.91953 |
| _246 | 0.555491 | 0.557654 | -0.00216 | | 0.002175 | 4.02469 |
| _037 | 0.54583 | 0.547862 | -0.00203 | | 0.00334 | 3.13054 |
| _455 | 0.511681 | 0.513016 | -0.00134 | | 0.001949 | 3.75511 |
| _446 | 0.5041 | 0.503887 | 0.000213 | | 0.003533 | 2.07878 |
| _067 | 0.450881 | 0.450392 | 0.000489 | | 0.004676 | 1.42246 |
| | | **AVERAGE** | **0.001589** | **0.002062** | **0.008252** | |

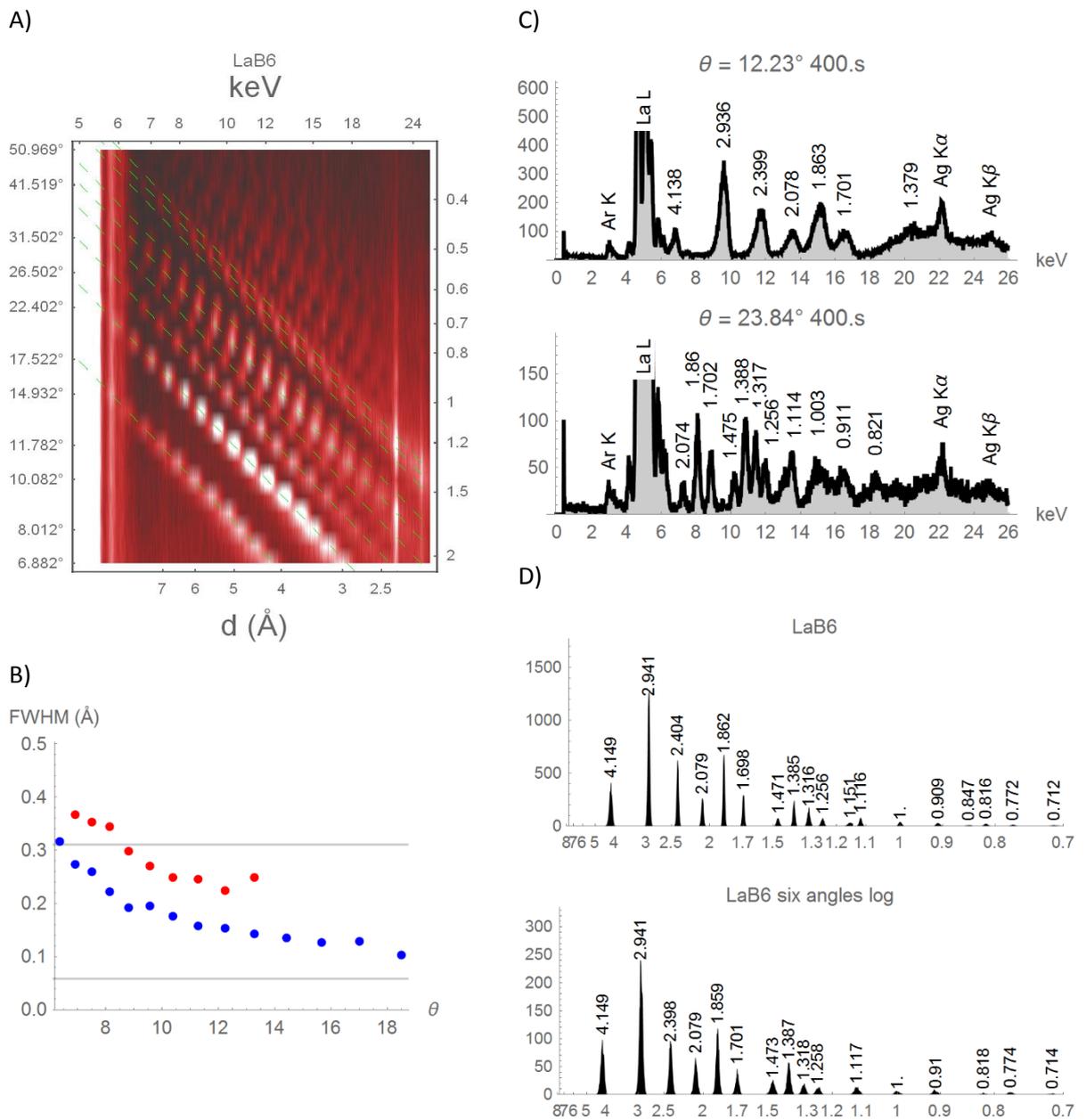

Figure 1: Plots for the analysis of coverage, precision and resolution of d-spacing detected on the standard LaB6 datasets. A) Log transformed density plot of the multiple angle scanning. B) Resolution (FWHM) of d-lines 2.93938 A (blue) and 4.15692 A (red) of the corresponding peaks measured at different angles. C) Diffractograms measured at angles 12.23° and 23.84°. D) Hybrid constructed from the 25 angles datasets and six angles datasets. The horizontal axes is spaced in the scale of q, labeled with the corresponding d-spacing values.

## 3.2 Mineral identification in green stone

Some green stone minerals found on polished surface of sculpted archaeological objects can be confused as jade. Additionally, jadeite and omphacite mineral are commonly associated in jadeite-jade minerals and are difficult to distinguish by conventional ADXRD. Due to the rare presence of jade in Earth, the distinction of these minerals is relevant for example to trace migration paths by the identification of jadeite-jade in pre-Hispanic objects [9, 10] or Neolithic axes from Europe [11].

ED diffractograms of Guatemala jade at 10.385° and 15.66° were chosen for comparison with data reported by two previous EDXRD portable XRD setups for cultural heritage [4, 7] (Figure 2 A and B). It is shown that d > 2.9 Å are not identified at 15.66° while d < 1.59 Å are not detected at 10.385°. Higher d values are more excited for angles lower than 10.385°. A wider range is spanned by our hybrid diffractogram constructed from the datasets measured at 22 angles (200 s/angle), and six angles (50s/angle), (Figure 2 C-D).

The non-invasive analysis of a very light green stone piece (a Guatemala Jade sample) identified jadeite as the main crystalline mineral phase from the hybrid diffractogram of the 22 angles dataset (Figure 2C) and detected iron in the XRF spectrum. The diversity of jadeite-jade stones in nature is reflected in the variety of d-spacing values sets reported in ADXRD references databases. In spite of that and the additional deviations introduced by different instrument and methods (angular, energy dispersive or hybrid XRD system), d-spacing values obtained by the ED and hybrid method in this work are in close agreement with those reported.

The most intense lines characteristic of jadeite (database_code_amcsd 0004610 : 2.9104 Å (100 %) and 2.4469 Å (46.24 %)) were identified in the Guatemala jade sample at 2.915 Å and 2.445 Å. In the peak at 2.9104 Å overlap the XRD lines, reported by ADXRD (reference amcsd 0004610: 2.9104 (100 %), 2.8590 (40.15 %), 2.8183 (17.15 %)). This broadening of the peaks is due to the limited energy resolution of detector and not because the method itself. Additional prominent lines are detected at 1.575Å, 1.481 Å and 1.37 Å (reference: 1.5758 Å (14.3 %), 1.4758 Å (12.23%), 1.3679 Å (3.03 %)). Probable presence of actinolite (database_code_amcsd 0001982, (2.1653 Å (29.75 %), 1.5794 Å (22.04 %), 1.3663 Å (11.27 %)) could be explained by the high intense lines at 2.165 Å and the contribution to the lines 1.575 Å and 1.37 Å. The low intensity lines detected at 1.852 Å, 1.208 Å and 1.071 Å not present in jadeite-omphacite stone (ex. sample "dark green jadeite") neither in omphacite sample could be distinctive lines of other minerals not identified. Values of d-spacing for jadeite and omphacite lower than 1.0948 Å are not reported in the ADXRD databases consulted, but are detected by the hybrid system in this sample. Confirmation of these lines could provide additional source of discriminatory information for the identification of jadeite. These results show that intermediate (1 - 3 Å) and higher (4 - 6 Å) d-spacing are well detected by the hybrid system. All the intense lines identified in Guatemala jadeite sample in the 22 angles hybrid diffractogram are also identified using six angles hybrid diffractogram (Figure 2D).

A)
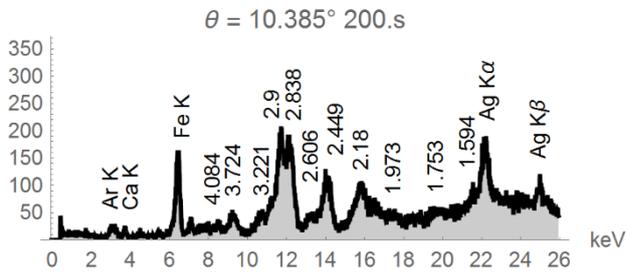

B)
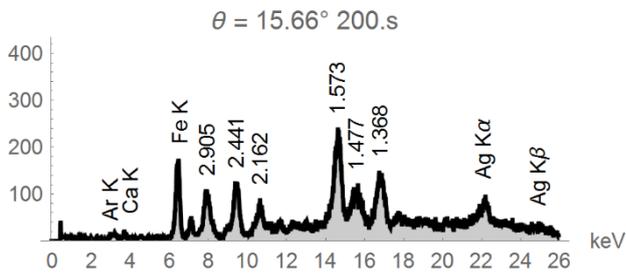

C)
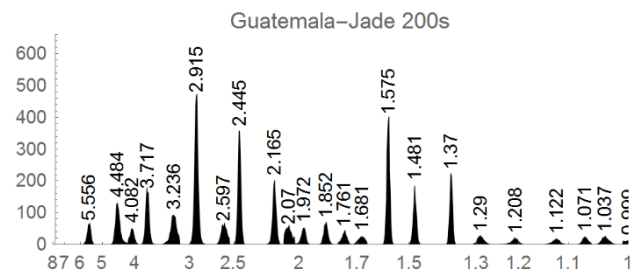

D)
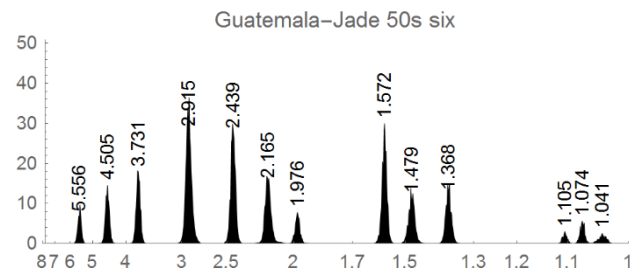

Figure 2 ED diffractograms of Guatemala jade sample at A) 10.385 ° and B) 15.66° and hybrid diffractograms of the C) 22 angles and D) six angles datasets in the range between 5° to 50°.

The six hybrid diffractograms of Guatemala jadeite (200s/angle), dark green jadeite and omphacite (Figure 3 A-C) are compared along the d-spacing range (> 1 Å), typically reported in ADXRD databases) and visibly differentiate . The detected d-spacing values are in close agreement with those reported by ADXRD database. Therefore, six scattering angles properly chosen can be sensible enough to accurately detect d-spacing values for the characterization and identification of jadeite.

A)

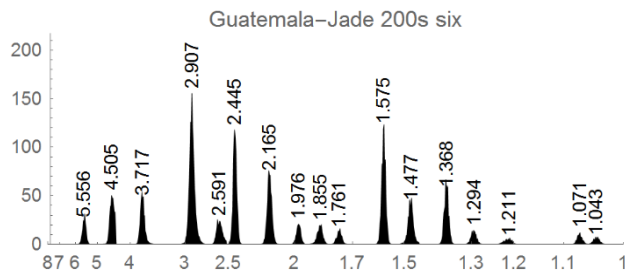

B)

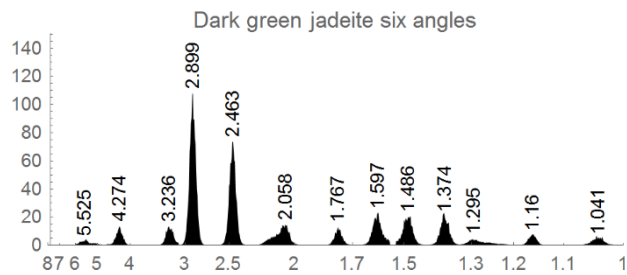

C)

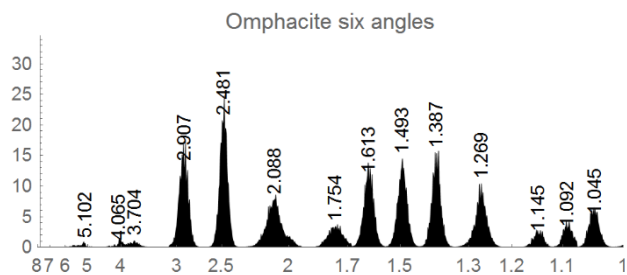

Figure 3 Six angle hybrid diffractograms of A) Guatemala jade, B) dark green jadeite and C) omphacite samples in the angular range between 5 to 50°.

Jadeite and omphacite are difficult to differentiate by XRD, therefore maximum coverage of XRD lines detected is then crucial for identification. The most intense lines in dark green jade sample (jadeite + omphacite) were detected at 2.899 Å and 2.463 Å which closely corresponds with jadeite (database_code_amcsd 0004603: 2.8941 Å (100 %) and 2.4663 Å (19.11 %)).  These lines (common to omphacite) further increase with the presence of omphacite. A line at 2.481 Å is detected with higher intensity in omphacite sample (database_code_amcsd 0005103: 2.8983 Å (38.91 %), 2.4784 Å (38.8 %)) with respect to dark green jadeite. Lines at 1.616 Å, 1.493 Å and 1.387 Å are more intense in omphacite sample.  The lines around 1.57 Å, 1.47 Å and 1.37Å are not prominent indicating that the sample is not pure jadeite or actinolite is not present. The presence of other mineral is further evidenced by the line at 1.616 Å (database_code_amcsd 000175: 1.6071 Å (15.14 %)) distinctive of omphacite. The reported relative intensities of d-spacing values are different between jadeite and omphacite. For quantitative analysis, the accurate determination of intensities require further developments and hybrid processing need to be evaluated for this purpose, and is not approached here. Lower intensities lines were detected in dark green jadeite sample and omphacite in the range 2 Å to 1 Å. Peaks positions for jadeite and omphacite differentiate by small shifts in this region.

The XRF spectra resulting from the same irradiated region in the sample complement and/or support the information obtained from XRD in hybrid diffractogram. For instance, the high presence of Ca in omphacite provide further information to distinguish purer jadeite sample (light green jade) from omphacite.

### 3.3 Pigment identification in the pictorial layer of painting

Paintings are challenging for analytical techniques because they are made of complex mixtures of materials distributed in layers [12]. Non-invasive analysis of paintings by portable X-ray diffraction apparatus (6) have been scarcely approached, even those cases where X-ray fluorescence alone does not bring sufficient information.

### 3. 3.1 Prussian blue

The hybrid AD/EDXRD-XRF analysis of modeled paintings allowed the identification of selected pigments usually not identified by XRF. For example, in the XRF spectrum of Prussian blue ($Fe_4[Fe(CN)_6]_3 \cdot xH_2O$), the organic component is not detected but high concentration of Fe is well identified. However, Fe is a very common element present also in dirty surface and colored under layer painted usually with earth pigments (iron based), and when artists decide a special composition of color in the exposed pictorial layer. Therefore, analysis by XRD can provide the complementary information required to identify the Prussian blue.

Besides Fe from Prussian blue, Ca, Ba and Zn are also detected in the analyzed modeled paintings by hybrid XRF spectra. The presence of Ca in painting spectra suggest calcite ($CaCO_3$) or gypsum ($CaSO_4$) use. The detection of Ba and Zn correspond to barite ($BaSO_4$) and zinc white ($ZnO$).

When no element responsible for color is identified by XRF and Fe is present, the choices to identify from the known blue pigments reduce to Prussian blue, Vivianite, silicate bases pigments (ex. ultramarine blue, smalt and Maya blue) or organic blues. Characteristic Prussian blue XRD lines well differentiate from ultramarine blue intense lines at 3.706 Å and 2.62 Å, according to d spacing reported for ultramarine pigment from Forbes Collection in the database of Museum of Fine Arts, Boston (references: PIG473).

In the hybrid diffractogram of the Prussian blue sample (pictorial layer on oil model painting canvas) (figure 4 a) the most intense d-spacing 5.076 Å, 3.597 Å, 2.538 Å and 2.288 Å were detected corresponding to PDF 01/073/0689 (at 5.08 Å (100 %), 3.59 Å (40 %), 2.54 Å (32%) and 2.27 Å (20%)). The lines 3.049 Å and 1.894 Å correspond to intense shifted lines of calcite PDF 04/007/8659 (at 3.03 Å (100 %), 1.91 Å (18%), 1.88 Å (20%)). Intense line of zinc white also contribute to 1.894 Å (1.91 Å (80%)). The lines at 2.538 Å and 2.288 Å overlap with calcite lines 2.49 Å (13%), 2.28 Å (19%). In the d spacing range from 2.079 Å to 1 Å there are interference of intensities lines of Prussian blue, calcite as well as the other inert materials present in the underlayers: barite (ex. 2.1 Å (55.66%)) and zinc white. Shifted intense lines of barite (PDF 04/007/7651) (3.43 Å (100 %), 3.31 Å (66.07%), 3.09 Å (98.20%)) and zinc white (3.16 Å (100 %), 3.12 (70 %), 3.04 (71%)) enhance the lines at 3.597 Å and 3.049 Å.

The pure Prussian blue pigment (high tinting strength) is often mixed in commercial blue oil paints with various inert or filler (i.e.: calcite, barite, dolomite, gypsum, kaolinite), but information of their chemical composition is usually not provided. In the hybrid XRD analysis of our Prussian blue painting (pure), the most intense line at 3.049 Å should correspond to thick preparation layer of calcite or gypsum (3.0754 Å). The absence of lines at 7.5(6) Å and 4.26(30) Å (Museum of Fine Arts reference: PIGCASO4 Gypsum and database_code_amcsd 0001807) exclude gypsum. Shifted lines coming from preparation or pictorial underlayers support the presence of specific inert filling material for preparations of not professional modern pigment.

In the hybrid diffractogram of Prussian blue acquired at six angles the most intense lines are identified. and d-values are very close to those obtained in the hybrid diffractogram constructed using 24 angles, subscribing the more economic choice of six angle to identify the mineral (Figure 4 b). Such experimental configuration (six angles) are particularly convinient when great amount of samples need to be measured, as is the case of painting attribution and authentication. However, angle scanning have be setup to detect the d spacing of 5.08 Å, avoiding XRF interferences. From the other side, the others XRD lines of Prussian blue can be interfered by XRD lines of inert materials.

Prussian blue can be identified by ADXRD by its strong diffraction line d= 5.08 Å detected at $2\theta=17.55°$ with a Cu anode tube. This is possible when the geometry of the XRD is carefully controlled. This angle is close to the usual lower limit reached in reflection XRD by portable XRD [13], and an accurate object positioning (sample in the center of Ewald sphere) procedure is crucial. On the other hand, an laboratory (AD)XRD study of Prussian blue in modeled paintings containing a mixture of Prussian blue $Fe_4[Fe(CN)_6]_3 \cdot 14H_2O$ and $BaSO_4$, a common filler, reported that pigment was not detectable if its concentration is lower than 2% [14]. Hybrid AD/EDXRD system can help to overcome this issue avoiding the geometrical limit.

In hybrid system, the line 5.08 Å can be detected at different angles (see the Angle vs. Energy densiplot in Figure 4) and hybrid system can profit for the best selection of angles to enhance the identification. In this sense, we found a curious result during the analysis of original EDXRD diffractograms: a narrow and strong peak correspond to 5.08 Å was detected at θ=23.84°, at the energy of 2.98 keV (Ar kα).

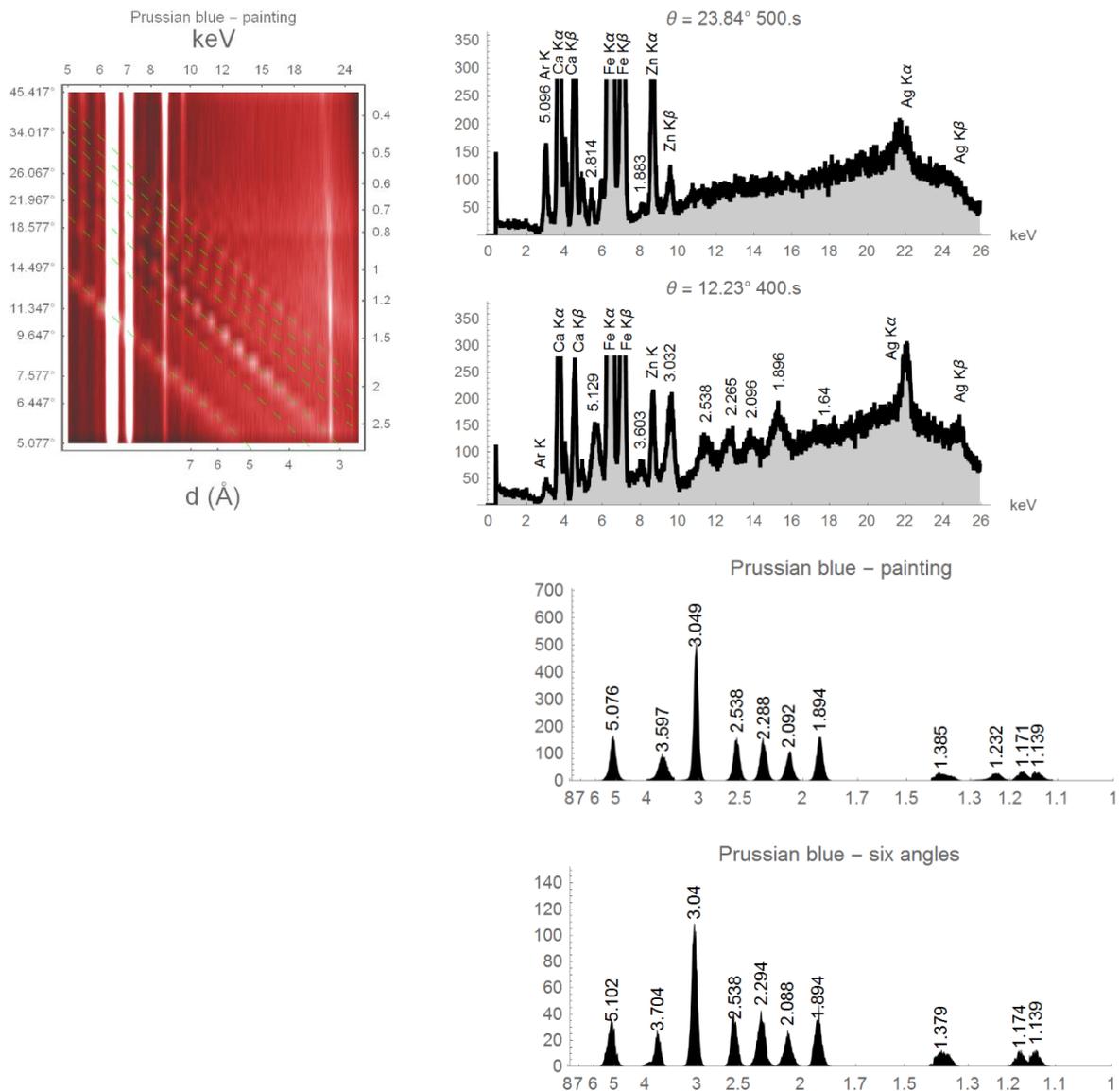

Figure 4 Log density Angle vs Energy density plot (top left), ED diffractograms at 12.23° and 23.84° and Hybrid diffractograms of Prussian blue painting taken with 24 in the angular range between 5 to 50°.

It can be explained as an enhancement of Ar XRF line due to the excitation of Ar present in the sample-detector path by XRD photons with energy higher and near to the absorption edge of Ar. The last can be confirmed when EDXRD around 23.84° are compared in the log E vs Angle density plot and with respect to other EDXRD at different angle i.e. 12.23° (Figure 4 b and c). ED diffractogram at 23.84° could be a very convenient choice to confirm the presence of Prussian blue for in situ practice since measurement at low angle are not always possible. Object's geometry and/or physical dimensions of instrument main components (X-ray tube and/or detectors containers) can obstruct the selection of lower angles.

### 3.4 Acquisition time

In EDXRD, measurement at a single angle produce a diffractogram spanning a d-spacing range at faster acquisition time than ADXRD. In ADXRD, a densely sampled angular range scanning is required to obtain a full diffractogram. In the hybrid AD/EDXRD system, EDXRD measured at multiple angles at a coarser sampling step can sweep the same range of d-spacing. For instance, six angles distributed with the proposed logarithmic scale [8], required a total acquisition time of 40 minutes (6 x 400s (time/angle)) for the standard $LaB_6$, a significant reduction compared with a dense angular scanning with equidistance angular step of 0.25° that can last around 9 hours [8], similar to ADXRD portable system without X-ray optics and/or 2D detector. The average accuracy of d-spacing obtained for this standard was 0.002062 Å.

A total acquisition time of 20 min using the hybrid six angle EDXRD scanning (200 s/angle) was enough to detect all interesting lines of jadeite for applications of mineral identification in green stone. No appreciable deterioration in accuracy and resolution was obtained by 50sec/angle measurements requiring a total acquisition time of only 5 minutes, detecting the more interesting lines of jadeite (see figure 2). It should be emphasized that this acquisition time is comparable to those reported using ADXRD portable system equipped with X-ray optics for quartz [3].

The acquisition time of six angles scanning for pigment identification in paintings, exemplified by the Prussian blue took 40 minutes (400 s/angle) or 20 minutes (200 s/angle) to be identified in the modeled paintings. This time is comparable (or even lower) with the fastest reported paintings analysis using portable ADXRD system [3]. Acquisition time can be optimized in correspondence to the intensity delivered at each angle, to obtain comparable signal to noise per angle.

### 4. CONCLUSIONS

The suitability for non-invasive surface mineral identification of the hybrid low power angle/energy dispersive X-ray diffraction and fluorescence portable system is evaluated. Performance in terms of lines, accuracy and time is assessed and compared with other portable systems. The hybrid EDXRD/XRF analysis using multiple angle scanning of standard (for line position and shape line calibration) and samples shows improvement of d-spacing accuracy, covering the full interesting d spacing range, with respect to EDXRD taken at single angle, and the convenience to reduce total acquisition time of hybrid measurements without appreciable degradation of signal. So far, time acquisition of 5 min to 20 min are reached using an economic and compact setup, which is lower or comparable with those times reported in similar applications with the fastest ADXRD based portable XRD/XRF instruments equipped with X-ray optics and/or 2D detectors. Further, even when lines lower than 1 Å are not reported in reference databases for our present study cases, strong enough lines tracing a clear Bragg path are detected in this range, promising an additional source of discriminatory information.

The analysis of samples of archaeometrical interest point to the potential use of this hybrid system in the study of archaeological and artworks objects. The non-destructive hybrid XRD/XRF analysis allowed identification of mineral in green stone: distinction between jadeite and omphacite. The identification of pigments, such as Prussian blue (typically not identified by Portable XRF or Portable ADXRD systems), in surface pictorial layer of modeled paintings was successfully carried, differentiating their corresponding more intense XRD lines from underlayes materials lines. Angular scanning efficient design along a minimal number of angles shown to be an effective cost/benefit choice especially appropriate in scenarios where a large number of sample need to be analyzed, as in the cases of migration paths or provenance studies. Hence, the hybrid system turns out particularly appropriate for the non-invasive analysis of artwork objects and for archaeometrical studies.

**Acknowledgements**

The data was measured in the Multidisciplinary Laboratory facilities at the International Center for Theoretical Physics (ICTP). We are grateful to Prof. Dr. John Niemela, Head of Multidisciplinary Laboratory, for encouraging this research, allowing the authors to carry out these measurements. The methods and software for processing the hybrid XRD-XRF data were developed at the Havana´s Archaeometry Laboratory with the auspicious of Havana Historian´s Office.